\documentclass[10pt,twocolumn,a4paper]{IEEEtran}
\usepackage{amsmath}
\usepackage{graphicx}

\newcommand{\figura}[3]
{
\begin{figure}
  \centering
 \includegraphics[width=7cm]{#1}
  \caption{#2}\label{#3}
\end{figure}
}
\title{On The Linear Behaviour of the Throughput of IEEE 802.11 DCF in Non-Saturated Conditions}
\author{\authorblockN{F. Daneshgaran, M. Laddomada, F. Mesiti, and M.
Mondin}
\thanks{F. Daneshgaran is with ECE Dept., CSU,
Los Angeles, USA.}
\thanks{M. Laddomada, F. Mesiti, and M. Mondin are with DELEN, Politecnico di Torino,
Italy.}}
\begin{document}
\maketitle
\begin{abstract}
We propose a linear model of the throughput of the IEEE 802.11
Distributed Coordination Function (DCF) protocol at the data link
layer in non-saturated traffic conditions. We show that the
throughput is a linear function of the packet arrival rate (PAR)
$\lambda$ with a slope depending on both the number of contending
stations and the average payload length. We also derive the
interval of validity of the proposed model by showing the presence
of a critical $\lambda$, above which the station begins operating
in saturated traffic conditions.

The analysis is based on the multi-dimensional Markovian state
transition model proposed by Liaw \textit{et al.} with the aim of
describing the behaviour of the MAC layer in unsaturated traffic
conditions. Simulation results closely match the theoretical
derivations, confirming the effectiveness of the proposed linear
model.
\end{abstract}
\section{Introduction}
Modelling of the DCF at the MAC layer of the series of IEEE 802.11
standards has recently garnered interest in the scientific
community~\cite{Bianchi}-\cite{WLee}. After the seminal work by
Bianchi~\cite{Bianchi} who proposed a bi-dimensional Markov model
of the back-off stage procedure adopted by the DCF in saturated
conditions, many papers have focused on various facets of basic
access mechanism providing extensions to most recent versions of
the IEEE 802.11 series of standards~\cite{standard_DCF_MAC}.
Recently, in \cite{Kumar} the authors proposed a novel fixed-point
analysis of the DCF providing an effective framework for analyzing
single cell IEEE 802.11 WLANs without resorting to the
bi-dimensional contention model~\cite{Bianchi}.

Practical networks usually operate in non-saturated conditions and
data traffic is mainly bursty. Under these operating conditions,
Bianchi's model does not describe accurately the behaviour of the
throughput at the MAC layer. In this respect,
in~\cite{Liaw}-\cite{hamilton} the authors proposed two different
bi-dimensional Markov models accounting for unsaturated traffic
conditions, extending the basic bi-dimensional model proposed
in~\cite{Bianchi}.

In this paper we take a different approach with respect to
works~\cite{Liaw}-\cite{hamilton}. Upon starting from the
bi-dimensional model proposed by Liaw \textit{et al.}
in~\cite{Liaw}, we show that the behaviour of the throughput of
the IEEE 802.11 DCF in unsaturated conditions can be described by
a linear relation that, with respect to the PAR $\lambda$, depends
on two network parameters: the number $N$ of contending stations,
and the average size $E[PL]$ of the transmitted packets. This is
one of the key contribution of the paper: no simulations are
needed for throughput evaluation since it can be theoretically
predicted employing the model $S(\lambda)=N\cdot E[PL]\lambda$
developed in Section~III. Of course, the limit of validity of such
a model has to be clearly identified, and it represents another
contribution of this paper. To this end, we derive the interval of
validity of the proposed model with respect to the PARs at the MAC
layer. We demonstrate the existence of a critical PAR,
$\lambda_c$, which discriminates the unsaturated region,
characterized by the range $\lambda\in [0,\lambda_c)$, from the
saturation zone identified by any $\lambda\in
[\lambda_c,+\infty)$.

For conciseness, we invite the interested reader to refer
to~\cite{Liaw} for many details on the considered bi-dimensional
Markov model, and references therein to get a picture of the topic
addressed in this letter. Briefly, Liaw \textit{et al.} extended
the saturated Bianchi's model by introducing a new idle state, not
present in the original Bianchi's model, accounting for the case
in which the station buffer is empty, after a successful
completion of a packet transmission. The main advantages of such a
model rely on its simplicity and the effectiveness in describing
the dynamics of the DCF in unsaturated traffic conditions, while
basic hypotheses are the same as in Bianchi's model.

Paper outline is as follows. In section~II, we briefly recall the
main probabilities needed for developing the proposed linear
model, evaluate the throughput and present the adopted traffic
model. Finally, Section~III presents the linear model of the
throughput along with simulation results.
\section{Problem formulation and Traffic model}
The bi-dimensional contention Markov model proposed in~\cite{Liaw}
governs the behaviour of each contending station through a series
of states indexed by the pair $(i,k),~\forall i\in [0,m], k\in
[0,W_i-1],$ whereby $i$ identifies the backoff stage, and $k\in
[0,W_i-1]$ the backoff counter. The other parameters needed in the
proposed framework can be summarized as follows: $\tau$ is the
probability that a station starts a transmission in a randomly
chosen slot time (ST), $q$ is the probability that there is at
least a packet in the queue after a successful transmission,
$W_i=2^i W_0,~\forall i\in [1,m],$ is the size of the $i$th
contention window, $W_0$ is the minimum size of the contention
window, $P_{I,0}$ is the probability of having at least one packet
to be transmitted in the queue when the system is in idle state,
and $p$ is the collision probability defined as in~\cite{Bianchi}
\begin{equation}\small
\label{eq:pcoll} p = 1 - (1-\tau)^{N-1}
\end{equation}
Stationary probability $b_I$ of being in the idle state is:
\begin{equation}\small
\label{eq:b_N} b_I  =  (1-q) b_{0,0}/P_{I,0}
\end{equation}
whereby $b_{0,0}$ is defined as follows:
\begin{equation}\small
\label{eq:b00} b_{0,0} =\frac{1-b_I}{\alpha},~\alpha = \frac{1}{2}
\left[ \frac{1-(2p)^{m+1}}{1-2p} W_0 + \frac{1-p^{m+1}}{1-p}
\right]
\end{equation}
By employing the normalization condition~\cite{Bianchi}, 
it is possible to obtain:
\begin{equation}\small
\label{eq:tau}
\tau  =  \sum_{i=0}^{m} b_{i,0}=\sum_{i=0}^{m} p^i \cdot b_{0,0}=\varepsilon \cdot b_{0,0},~\varepsilon =\frac{1-p^{m+1}}{1-p} 
\end{equation}
Next line of pursuit is the computation of the system throughput.
Putting together Eq.s~(\ref{eq:pcoll}),~(\ref{eq:tau}), a
nonlinear system can be defined and solved numerically, obtaining
the values of $\tau$ and $p$. The solution of the previous system
is used for evaluating the throughput, defined as the ratio
between the average payload information transmitted in a ST and
the average length, $T_{av}$, of a ST:
\begin{eqnarray}\small
\label{eq.system2} S   & = &P_t \cdot P_s \cdot
E[PL]/T_{av}
\end{eqnarray}
whereby $E[PL]$ is the average packet payload length (expressed in
bits), $P_t$ is the probability that there is at least one
transmission in the considered ST, with $N$ stations contending
for the channel, each transmitting with probability $\tau$, i.e.,
$P_t = 1 - (1 - \tau) ^ N$. Probability $P_s$ is the conditional
probability that a packet transmission occurring on the channel is
successful:
\begin{equation}\small
\label{equat_PS} P_s = N\tau(1-\tau)^{N-1}/P_t
\end{equation}
Upon noting that, in a given ST, a station can reside in one of
three possible kind of states, namely the idle state $I$ where the
station spends $T_I$, the backoff states where the station spends
$T_{BO}$, and the transmitting states in which the station spends
$T_{TX}$, then the average duration $T_{av}$ of a ST easily
follows
\begin{eqnarray}\small
\label{eq:tAV}
T_{av} & = & b_I \cdot T_{I} + \tau \cdot T_{TX} + \sum_{i=0}^{m}\sum_{k=1}^{W_i-1} b_{i,k} \cdot T_{BO} \nonumber \\
       & = & b_I \cdot T_{I} + [\varepsilon \cdot T_{TX} + \theta \cdot T_{BO}] \cdot b_{0,0} 
\end{eqnarray}
where $\varepsilon$ is defined in (\ref{eq:tau}), and
\begin{equation}\small
\label{eq:theta} \theta = \frac{1}{2} \left[
\frac{1-(2p)^{m+1}}{1-2p} W_0 - \frac{1-p^{m+1}}{1-p} \right]
\end{equation}
Let us define the time durations $T_{TX},~T_{BO}$ and $T_I$
in~(\ref{eq:tAV}). Transmission time $T_{TX}$ can be evaluated by
noting that a station can experience two possible events: it
successfully transmits over the channel or it encounters a
collision. By doing so, $T_{TX}$ can be defined as follows:
\begin{eqnarray}\small
\label{tTX} T_{TX} & = & (1-p) \cdot T_{s} + p \cdot T_{c}
\end{eqnarray}
whereby $T_c$ and $T_{s}$ are, respectively, the average time a
channel is sensed busy due to a collision, and the successful data
frame transmission time \cite{Bianchi,Liaw}.

Backoff time duration $T_{BO}$ can be evaluated by considering the
following two possibilities. A station can reside in a backoff
slot of duration $\sigma$ if no other station is transmitting in
the same ST, or for a time $T_{TX}$ due to a collision with at
least another station occupying the channel:
\begin{equation}\small
  \label{tBO}
  T_{BO} = (1 - P_{tx[N-1]}) \cdot \sigma + P_{tx[N-1]} \cdot T_{TX}
\end{equation}
whereby $P_{tx[N-1]} = 1 - (1 - \tau)^{N-1}$ corresponds to the
probability that at least a station, other than the tagged one, is
transmitting in a ST.

For the sake of defining the traffic model employed for
performance verification, we need to define both the access time
$T_A$ (this is the average time a station spends through the
various backoff stages before transmitting a packet) and the
service time $T_S$. From~\cite{Liaw}, $T_A$ can be defined as
follows:
\begin{equation}\small
\label{tA}
T_{A} = \frac{\sum_{i=0}^{m}p^i \cdot \frac{W_i}{2} \cdot T_{BO}}{\sum_{i=0}^{m}p^i}=\frac{  W_0}{2\, \varepsilon}\frac{1-(2p)^{m+1}}{1-(2p)} \cdot T_{BO} 
\end{equation}
whereby $\varepsilon$ is as defined in (\ref{eq:tau}). On the
other hand, $T_S$~\cite{WLee}, i.e., the time elapsed from the
moment a packet is taken from the queue to the instant in which it
is successfully transmitted, can be defined as $T_S=  T_{A} +
T_{TX}$.

As far as $T_I$ is concerned, we resort to the
definition~\cite{Liaw}:
\begin{equation}\small
\label{eq:tI_final} T_{I}  =  \frac{[\varepsilon \cdot T_{TX} +
\theta \cdot T_{BO}] \cdot b_{0,0}}{1-b_I} =\frac{[\varepsilon
\cdot T_{TX} + \theta \cdot T_{BO}]}{\alpha}
\end{equation}
where last equality stems from~(\ref{eq:b00}).

The employed traffic model is $M/G/1/K$. Probabilities $q$ and
$P_{I,0}$ in our model can be defined as follows:
\begin{equation}\small
\label{eq:q}
\begin{array}{rlrl}
q   = & 1-\pi_0= 1 - \frac{1-\rho}{1-\rho^{K+1}}, & P_{I,0}= & 1 -
e^{-\lambda \cdot T_{I}}
\end{array}
\end{equation}
where $\pi_0$ is the probability of an empty system~\cite{bolch},
$\rho= \lambda\cdot T_S$, $q$ follows from M/G/1/K queuing
theory~\cite{bolch}, while $P_{I0}$ stems from the fact that for
exponentially distributed interarrival times with mean $1/\lambda$
the probability of having at least one packet arrival during time
$T$ is equal to $1 - e^{-\lambda \cdot T}$.

Employing~(\ref{eq:b_N}), for a $K-1$-length queue, we have:
\begin{equation}\small
\label{eq:bI_traffic} b_I = \frac{1-\rho}{1-\rho^{K+1}} \cdot
\frac{1}{1 - e^{-\lambda \cdot T_{I}}} \cdot b_{0,0}
\end{equation}
\section{The Linear model and simulation results}
\begin{table}\caption{Numerical Results.}
\begin{center}
\begin{tabular}{l||c|c|c}\hline

\hline\hline $N$ &      10 & 20 & 30 \\\hline

$S_m$ [Mbps]&            9.118& 8.73& 8.608\\\hline

$\lambda_c$ [pkt/s]&    111.2&53.235&34.99\\\hline

\hline\hline
\end{tabular}
 \label{tab.lambdacritici}
\end{center}
\end{table}
A model of the throughput in non-saturated traffic conditions,
along with its dependence on some key network parameters, can be
derived by analyzing~(\ref{eq.system2}) in the limit
$\lambda\rightarrow 0$.

Let us write the throughput in~(\ref{eq.system2}) as a function of
$\tau$. By employing~(\ref{equat_PS}), the numerator can be
rewritten as $ P_t P_s E[PL]=N\tau (1-\tau)^{N-1}E[PL]$. As far as
the denominator is concerned, upon
substituting~(\ref{eq:tI_final}) in~(\ref{eq:tAV}), and
remembering that $b_{0,0}=(1-b_I)/\alpha$, after some algebra, it
is possible to obtain the following relation $ T_{av}=[\varepsilon
\cdot T_{TX} + \theta \cdot T_{BO}]/\alpha$. By collecting the
previous relations, the throughput can be rewritten as:
\begin{equation}\small\label{throughput_S}
S(\tau)=\frac{N\tau (1-\tau)^{N-1}E[PL]
\alpha(\tau)}{\varepsilon(\tau) \cdot T_{TX}(\tau) + \theta(\tau)
\cdot T_{BO}(\tau)}
\end{equation}
whereby we highlighted the dependence on $\tau$ of the terms
$\alpha,~\varepsilon,~T_{TX},~\theta$, and $T_{BO}$. Upon noting
that~(\ref{eq:b00}) yields $\lim_{\lambda\rightarrow 0}b_{0,0}=
0$, from~(\ref{eq:tau}) it follows $\tau\rightarrow 0$ as well.
In the limit $\tau\rightarrow 0$, it is straightforward to
demonstrate the following relations: $p\rightarrow 0$
(from~(\ref{eq:pcoll})), $\varepsilon(\tau)\rightarrow 1$
(from~(\ref{eq:tau})), $\theta(\tau)\rightarrow \frac{W_0-1}{2}$
(from~(\ref{eq:theta})), $\alpha(\tau)\rightarrow \frac{W_0+1}{2}$
(from~(\ref{eq:b00})), $T_{TX}(\tau)\rightarrow T_s$
(from~(\ref{tTX})), and $T_{BO}(\tau)\rightarrow \sigma$
(from~(\ref{tBO})).

Upon substituting the derivations above in~(\ref{throughput_S}),
the limit $\lim_{\tau\rightarrow
0}S(\tau)=0=\left[S(\tau)\mid_{\tau=0}\right]$ easily follows.

Upon employing the Taylor's formula around $\tau\approx 0$, the
throughput $S(\tau)$ can be well approximated as follows:
\[
S(\tau) \approx
\left[S(\tau)\mid_{\tau=0}\right]+\left[\frac{\partial
S(\tau)}{\partial \tau}\mid_{\tau=0}\right]\tau
=\left[\frac{\partial S(\tau)}{\partial
\tau}\mid_{\tau=0}\right]\tau
\]
As far as $\tau\approx 0$, $(1-\tau)^{N-1}\approx
1-(N-1)\tau+o(\tau)$; therefore, the approximation
$N\tau(1-\tau)^{N-1}\approx N\tau +o(\tau)$ holds as well. As a
consequence,~(\ref{throughput_S}) can be rewritten as follows:
\begin{table}\caption{Typical network parameters}
\begin{center}
\begin{tabular}{c|c||c|c}\hline
\hline MAC header & 28 bytes&Propag. delay $\tau_p$ & 1 $\mu s$\\
\hline PLCP Preamble & 144 bit & PLCP Header & 48 bit \\
\hline PHY header & 24 bytes&Slot time & 20 $\mu s$\\
\hline PLCP rate & 1Mbps & W$_0$ & 32\\
\hline No. back-off stages, m & 5 & W$_{max}$ & 1024\\
\hline Payload size & 1025 bytes&SIFS & 10 $\mu s$\\
\hline ACK & 14 bytes&DIFS & 50 $\mu s$\\
\hline ACK timeout & 364$\mu s$ &EIFS & 364 $\mu s$\\
%

\hline\hline
\end{tabular}
 \label{tab.design.times}
\end{center}
\end{table}
\begin{equation}\small
\label{throughput_S_linear}
S(\tau)\approx \frac{N\cdot E[PL]\cdot
\frac{W_0+1}{2}}{T_s+\frac{W_0-1}{2}\cdot\sigma}\cdot \tau
\end{equation}
Next line of pursuit consists in expressing $\tau$ in terms of the
PAR $\lambda$. By using the MacLaurin expansion of the exponential
$e^{-\lambda T_I}\approx 1-\lambda T_I+o(\lambda)$,
from~(\ref{eq:bI_traffic}) it is $b_{0,0}\approx \lambda T_I$ as
$\lambda\rightarrow 0$. On the other hand,
Equ.~(\ref{eq:tI_final}) yields $\lim_{\lambda\rightarrow
0}T_I=[T_s+\frac{W_0-1}{2}\sigma]/\frac{W_0+1}{2}$. Finally, in
the limit $\lambda\rightarrow 0$, Equ.~(\ref{eq:tau}) yields $\tau
\rightarrow b_{0,0}\approx \lambda\cdot
[T_s+\frac{W_0-1}{2}\sigma]/\frac{W_0+1}{2}$. Upon substituting
the previous mathematical derivations in
(\ref{throughput_S_linear}), the throughput can be approximated as
follows:
\begin{equation}\small
\label{lin_model_throughput}
S(\lambda)\simeq N\cdot E[PL]\cdot\lambda
\end{equation}
It is interesting to estimate the interval of validity
$[0,\lambda_c]$ of the linear throughput model proposed
in~(\ref{lin_model_throughput}). An appropriate value of
$\lambda_c$ can be obtained by finding the abscissa $\lambda$
corresponding to the intersection of the straight line
(\ref{lin_model_throughput}) with the horizontal line passing
through the maximum $S_m$ of the throughput. To this end, after
expressing any term involved in~(\ref{eq.system2}) as a function
of $\tau$, the maximum throughput $S_m$ can be obtained in two
steps. First, one has to find the value $\tau_m$ for which the
throughput gets maximized. This can be easily obtained by equating
to zero the derivative of~(\ref{eq.system2}) with respect to
$\tau$, and then solving for $\tau$. In the second step, the
maximizing value $\tau_m$ can be substituted into $S(\tau)$ for
obtaining $S_m=S\left(\tau\right)_{|_{\tau_m}}$. Finally,
$\lambda_c$ is the value of $\lambda$ for which the following
holds:
\figura{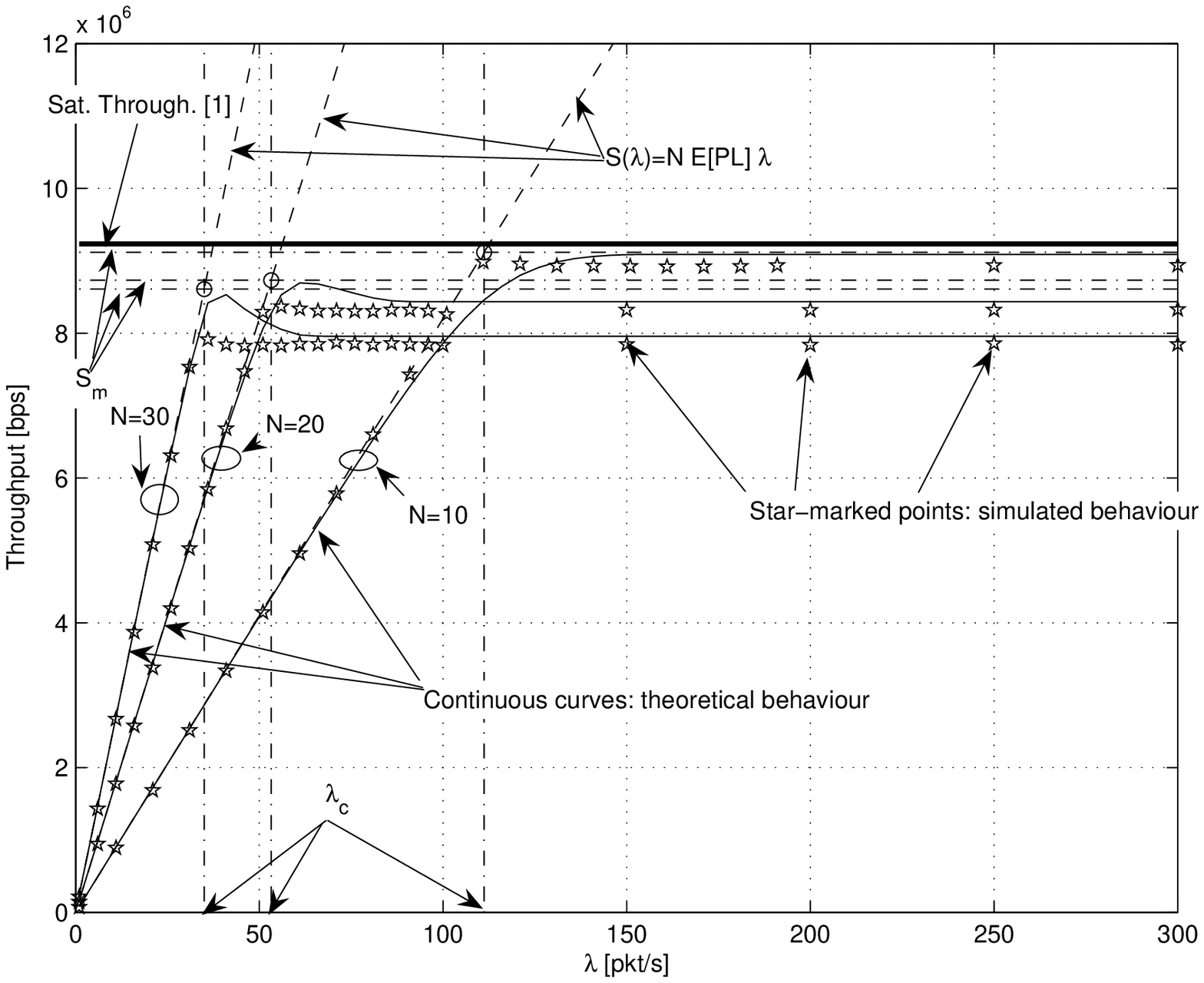}{Throughput for the 2-way
mechanism as a function of the PAR $\lambda$, for three different
number $N$ of contending stations. Straight dashed lines refer to
the linear model of the throughput
in~(\ref{lin_model_throughput}).}{linearity_throughput_lambda}
\begin{equation}\small
\label{lambda_critici}
  \lambda_c  = \lambda : N\cdot E[PL]\cdot\lambda =S_m
\end{equation}
Table~\ref{tab.lambdacritici} shows the values of $\lambda_c$,
found numerically, along with the respective values of $S_m$, for
various values of $N$. We considered the bit rate 54Mbps for the
protocol IEEE 802.11g. The linear
model~(\ref{lin_model_throughput}) shows a close agreement with
both theoretical (continuous curves) and simulated throughput
curves. Fig.~\ref{linearity_throughput_lambda} shows the straight
lines in~(\ref{lin_model_throughput}) (dashed lines) for three
different values of $N$. The figure also shows the values of $S_m$
(horizontal dash-dot lines) along with the three values of
$\lambda_c$ deduced from~(\ref{lambda_critici}) and noted in
Table~\ref{tab.lambdacritici}. Simulated values (star-marked
points) have been obtained with ns-2 by using settings noted in
Table~\ref{tab.design.times}, along with standard 54Mbps 802.11
parameterizations~\cite{standard_DCF_MAC}, which are also the
standard parameters defined in ns-2. Within ns-2, $N$ stations
have been randomly placed in a square area with edge size equal to
$50$m using a uniform distribution. All simulation results in
Fig.~\ref{linearity_throughput_lambda} are obtained with a 95\%
confidence interval lower than $15$kbps.

Fig.~\ref{linearity_throughput_lambda} shows that the maximum
achievable throughput does not exceed 10Mbps (@N=10) despite the
maximum bit rate employed (54Mbps): this throughput penalty is
essentially due to the fact that the control packet and the PLCL
header are transmitted at 1Mbps no matter the operating
transmission mode. As a reference throughput performance, we show
the maximum saturation throughput (labelled "Sat. Through. [1]" in
Fig.~\ref{linearity_throughput_lambda}) found in \cite{Bianchi} in
case the $N$ contending stations transmit at the optimal
transmission probability $\tau$. Notice that, throughput penalty
with respect to 54Mbps is well predicted by the theoretical
formulation presented in \cite{Bianchi}.


\begin{thebibliography}{99}

\bibitem{Bianchi} G. Bianchi, "Performance analysis of the IEEE 802.11 distributed
coordination function'', {\em IEEE JSAC}, Vol.18, No.3, March
2000.

\bibitem{Bianchi2} G. Bianchi and I. Tinnirello, "Remarks on IEEE 802.11 DCF performance analysis'', {\em IEEE Comm. Letters}, Vol.9, No.8,
Aug. 2005.

\bibitem{Kumar} A. Kumar, E. Altman, D. Miorandi, and M. Goyal, "New insights from a fixed-point analysis of single
cell IEEE 802.11 WLANs'', {\em to appear on IEEE/ACM Trans. on
Networking}, 2007, now available on IEEExplorer.

\bibitem{Liaw}
Y.S. Liaw, A. Dadej, and A.Jayasuriya, "Performance analysis of
IEEE 802.11 DCF under limited load'', {\em In Proc. of
Asia-Pacific Conference on Communications}, Vol.1, pp.759 - 763,
03-05 Oct. 2005.

\bibitem{hamilton} D. Malone, K. Duffy, and D.J. Leith,
"Modeling the 802.11 distributed coordination function in
non-saturated heterogeneous conditions'', {\em IEEE-ACM Trans. on
Networking}, vol. 15, No. 1, pp. 159–172, Feb. 2007.

\bibitem{WLee}
W. Lee, C. Wang, and K. Sohraby, "On use of traditional M/G/1
model for IEEE 802.11 DCF in unsaturated traffic conditions'',
{\em In Proc. of IEEE WCNC}, 2006, pp. 1933–1937.

\bibitem{standard_DCF_MAC}{\em IEEE Standard for Wireless LAN Medium Access Control (MAC) and Physical Layer (PHY)
Specifications}, November 1997, P802.11

\bibitem{bolch} G. Bolch, S. Greiner, H. de Meer, and K.S. Trivedi, \textit{Queueing Networks and Markov Chains},
Wiley-Interscience, 2nd edition, 2006.

\end{thebibliography}
\end{document}